\documentclass[prb,amsmath,amssymb,preprint]{revtex4}

\usepackage{graphicx}
\usepackage{bm}

\newcommand{\figwidth}{3.85in}

\begin{document}

\title{The Duhem-Quine thesis and the dark matter problem}

\author{M. A. Reynolds}
\email{anthony.reynolds@erau.edu}
\affiliation{Department of Physical Sciences,
Embry-Riddle Aeronautical University,
Daytona Beach, Florida, 32114}

\date{\today}

\begin{abstract}
There are few opportunities in introductory physics for a genuine discussion of the
philosophy of science,
especially in cases where the physical principles are straightforward
and the mathematics is simple.
Terrestrial classical mechanics satisfies these requirements,
but students new to physics usually carry too many
incorrect or misleading preconceptions about the subject for it to be analyzed epistemologically.
The problem of dark matter,
and especially the physics of spiral galaxy velocity rotation curves,
is a straightforward application of Newton's laws of motion and gravitation,
and is just enough removed from everyday experience to be analyzed 
from a fresh perspective.
It is proposed to teach students about
important issues in the philosophy of physics,
including Bacon's induction,
Popper's falsifiability,
and the Duhem-Quine thesis,
all in light of the dark matter problem.
These issues can be discussed in an advanced classical mechanics course,
or, with limited simplification,
at the end of a first course in introductory mechanics.
The goal is for
students to understand at a deeper level how the physics community has arrived at the current state of knowledge.

\begin{center}
Submitted to the \textit{American Journal of Physics}
\end{center}

\end{abstract}

\pacs{Valid PACS appear here}
\maketitle

\begin{quote}
\textit{Epistemology without contact with science becomes an empty scheme. 
Science without epistemology is---in so far as it is thinkable at all---primitive and muddled.}
-- Albert Einstein\cite{Schilpp}
\end{quote}

\section{Introduction}

One feature of physics education today (and science education in general)
is the lack of serious discussion of epistemological issues,
especially in the intermediate courses that cover the ``modern'' physics topics of 
relativity and quantum mechanics.
Some would argue that such discussion is irrelevant to doing physics,
but I contend (along with Einstein) that,
regardless of whether or not it affects concrete results,
a knowledge of issues such as the ``problem of induction'' and
the ``problem of demarcation'' enrich our enjoyment and understanding of the physical universe.\cite{Howard05}
These issues are especially important today, when topics such as
string theory,
supersymmetry,
and the multiverse landscape 
dominate the popular physics literature,\cite{Baggott13}
and in which students show an intense interest.
While complex issues should probably wait until students have a working
knowledge of a wide variety of physical concepts,
I suggest that a conversation should be started in the introductory physics sequence,
and that a student's first taste of the philosophy of physics is more palatable
if the discussion centers around classical mechanics.
The problem of dark matter offers a practical demonstration that is ideal for this task.

In this article, I outline a presentation that is well-suited to a junior-level classical mechanics course
where the students have the requisite mathematical background,
but it can be easily simplified to fit at the end of a first physics course
after the standard coverage of Newton's laws of motion and gravitation and Kepler's laws of planetary motion.
After a brief survey of some of the philosophical issues in Sec.~\ref{sec:Duhem},
and a short account of the classical evidence for dark matter at the beginning of Sec.~\ref{sec:Dark},
a case study is presented of the dark matter problem focusing on the analysis of galactic rotation curves.
Among other suggestions,
a simple version of MOND
(MOdified Newtonian Dynamics)
is outlined,
and the study finishes with an attempt to reconcile the observations and theoretical predictions
in the light of the ideas of verification and falsification discussed in Sec.~\ref{sec:Duhem}.

\section{The Duhem-Quine Thesis\cite{Harding76}}
\label{sec:Duhem}

Is there a scientific method?
If there is, how do we distinguish it (or demarcate it) from other enterprises that are not scientific?
The traditional answer,
given by Francis Bacon in the 17th century,
is that science is inductive.
That is, scientists determine general ``laws of nature'' from many observations of 
specific instances or events (i.e., experiments),
and as more observations are made,
those laws are refined or perhaps changed drastically.\cite{Shapin96}
Other statements, while they may be general,
are not considered scientific
if they are not based on observations, but are based on authority or emotion, for example.

Since that time,
many philosophers
(most notably David Hume in the 18th century)
have recognized that this answer was logically inconsistent.
While deductive reasoning can prove the truth of a conclusion 
given that the premises are true,
this same rational justification does not apply to inductive reasoning.
For example,
the conclusion ``John is tall'' follows directly if the premises
``all men are tall'' and ``John is a man'' are true,
but no number of observations of tall men,
no matter how large,
can ever prove that ``all men are tall.''
And worse,
contrary to our common sense,
increasing the number confirming observations does not logically entail any increase in the probability of the truth of the conclusion.
Yet this is the essence of the Baconian method.
This ``problem of induction'' has puzzled philosophers ever since. 

In the 20th century, Karl Popper
pointed out the asymmetry in the way observations verify or falsify hypotheses.\cite{Popper59,Magee}
That is, hypotheses can be definitively falsified but not definitively verified.
No number of observations agreeing with a statement can verify it,
but a single observation disagreeing with that statement can falsify it.
Popper thus offered falsifiability as a solution to the problem of induction.
He claimed that induction is superfluous because science advances not by induction, but by attempts to falsify hypotheses.
It can also solve the problem of demarcation,
because scientific theories must consist of statements that are falsifiable.

Most physicists today don't agree with the naive view of falsification
(and neither did Popper)
as explaining the scientific method
because one can point to many episodes in the history of science that have not
followed the falsification process.
In addition,
there is another problem with naive falsification
that was pointed out by Pierre Duhem in 1906,\cite{Duhem}
namely that individual hypotheses cannot be falsified.
He claimed that
because each hypothesis is inextricably intertwined with so many auxiliary hypotheses and assumptions
that the physicist usually assumes to be beyond doubt,
any experiment that appears to falsify the main hypothesis under consideration
could in fact be falsifying one of the auxiliary hypotheses.
Such an experiment can show that one of the hypotheses is unsatisfactory,
but does not reveal which one.
As Duhem put it,
``[p]hysics is not a machine which lets itself be taken apart \ldots [but] a system that must be taken as a whole.''\cite{Duhemp8}
Furthermore,
Duhem thought that Bacon's ``crucial experiments,'' 
those that unambiguously determine which of two competing theories is correct,
are impossible.
Because of the auxiliary assumptions that any prediction rests on,
it is possible that a different set of assumptions might change the prediction to agree with the experimental result.
But,
``the physicist is never sure he has exhausted all the imaginable assumptions.''\cite{Duhemp11}
These ideas constitute the Duhem-Quine thesis.\cite{Quine}

Popper agreed with Duhem that isolated hypotheses are never tested,
and his suggestion was that physicists
``should \textit{guess} \ldots which part of such a system is responsible for the refutation,
perhaps helped by independent tests of some portions of the system.''\cite{Lakatos74}
As Popper put it,
``It is often only the scientific instinct of the investigator 
(influenced, of course, by the results of testing and retesting),
that makes him guess which statements \ldots he should regard as innocuous,
and which he should regard as being in need of modification.''\cite{Popper59p76}

Two famous examples of this in action were the discovery of the planet Neptune in 1846
and the anomalous perihelion advance of Mercury.
After the serendipitous discovery of Uranus in 1781,
the failure to correctly predict the position of Uranus
did not immediately cause astronomers to throw out the theory of Newtonian gravitation,
even though Newton's law of gravity was the prime hypothesis that allowed the 
perturbations due to other six planets to be calculated.
One reason was that Newtonian gravity had proved extremely successful in a wide variety of regimes,
from terrestrial to heavenly.
But neither was the assumption that there must be an eighth planet immediately championed.
By the 1840s the problem had not been solved,
and the two leading possible solutions were
``one, that the law of gravitation acts in some unexpected manner at
the enormous distance of Uranus, or two, \ldots that a planet lies beyond Uranus.''\cite{Smith89p398}
Interestingly,
there had been a third option put forward by Friedrich Wilhelm Bessel in the 1820s.\cite{Smith89}
His proposal entailed ``specific attraction,''
in which the gravitational mass of an object depended on its chemical composition,
and was not necessarily equal to its inertial mass.
However,
in experiments with pendula of different composition,
he found no evidence for this effect, and the idea was dropped.
Later, when a similar problem arose with the orbit of Mercury
--- its perihelion advanced by 43 arc seconds per century more than could be explained by planetary perturbations --- 
the opposite conclusion was found to be correct.\cite{Lahav14}
That is, 
the hypothesis that was modified was not the number of planets (the planet Vulcan was never found)
but Newton's law of gravitation (Einstein's general theory of relativity explained the anomaly).

These examples show that not only is it difficult to decide on which hypothesis to modify,
but, to echo Duhem,
one is never sure that all possible auxiliary assumptions have been accounted for.
Where does this leave physics and the scientific method?
Inductive inference is not rationally justifiable,
Popper's view of falsification means that none of our knowledge is certain,
and Duhem claims that crucial experiments cannot be performed.
However,
Harding points out that
recently philosophers have occupied a middle ground,
and contend that
``science definitely can be reconstructed as a rational enterprise,
but there may well be no principles for a prospective methodology of science.''\cite{Harding76pxx}
This echoes the feelings of many physicists who claim that physics is successful 
regardless of what philosophers have to say about its methods.

A concrete example that illustrates these ideas with a level of mathematics 
that is accessible to undergraduates is the dark matter problem.
Section \ref{sec:Dark}
introduces the problem and lays out a method of solution consistent with Duhem and Popper.
That is,
the hypotheses are clearly stated,
a prediction is made,
and several guesses are suggested as to which auxiliary hypotheses are in need of modification,
as well as how to modify them.

\section{The Dark Matter Problem}
\label{sec:Dark}

One of the first indications of the ``missing mass problem,'' as it was known then,
was the observation in 1933 of the Coma cluster cluster of galaxies by Fritz Zwicky.\cite{Zwicky33}
The Coma cluster is about 100 Mpc from Earth in the direction of the constellation Coma Berenices,
and
it consists of more than 1000 member galaxies within a radius of about 1 Mpc.
One parsec (pc) is about $3\times 10^{16}$ m.
Zwicky's measurement of the velocity dispersion of the member galaxies resulted in an estimate
for the total mass of the cluster.
Of course,
he made the assumption that the cluster was ``mechanically stationary,''
that is,
it was not in the process of flying apart.
Then he compared this result with another mass determination method,
that of inferring the mass of each individual member galaxy from its brightness
and adding them to obtain the total cluster mass.
This relied on a knowledge of the so-called mass-to-light ratio of galaxies,
which had been obtained from the velocity rotation curves of nearby galaxies, including the Milky Way.
Zwicky found that the cluster as a whole contained several hundred times more mass than expected
from the sum of the individual galaxy masses.
Today, this ratio has been reduced to about 30 
``using the best cluster and galaxy models,''\cite{BinneyTremaine}
which implies that most of the mass in the cluster is not visible.
Zwicky called this ``dark matter,''
but at that time he meant
``cool and cold stars, macroscopic and microscopic solid bodies, and gases,''\cite{Zwicky33p218}
not the exotic particles that are being searched for today.

The next step came in the early 1970s
when the velocity rotation curves of spiral galaxies were able to be measured out to large distances 
due to improved observational techniques.
Using dynamical techniques similar to Zwicky's,
and outlined below in Secs.~\ref{sec:Prediction} and \ref{sec:Observation},
Vera Rubin and coworkers\cite{Rubin70,Rubin78,Rubin80} measured the 
rotational velocity of objects far from a galaxy's center,
which in principle allowed a determination of its mass.
The result was, again,
that there was far more mass acting gravitationally than was visible in the form of stars and nebulae,
and that this invisible mass extended far beyond the visible limits of the galaxy.
These ``galactic rotation curves'' were the primary evidence for dark matter for several decades.

Now there are several other, more sophisticated methods.
For example, one technique is to observe the 
general relativistic effect of gravitational lensing
as distant light ``bends'' around a foreground galaxy.
Einstein, of course, had predicted the existence of so-called ``Einstein rings'' in 1912,\cite{Renn97}
and many have now been found.
Another, more recent technique is an interpretation of the high-$\ell$ shape of the
cosmic microwave background spectrum in light of the existence of cosmological dark matter.\cite{Skordis09}
However,
while any solution must match all of these observations,
for pedagogical purposes,
and the cleanest argument concerning the scientific method,
the focus here will be on the velocity rotation curves of spiral galaxies.
In addition,
even though general relativity is our current best theory of gravity
and any solution to the cosmological dark matter problem must be consistent with it,
the analysis will be restricted to Newtonian gravity to make the discussion mathematically accessible to undergraduates.

\subsection{The Hypothesis}

\begin{figure}
\includegraphics[width=\figwidth]{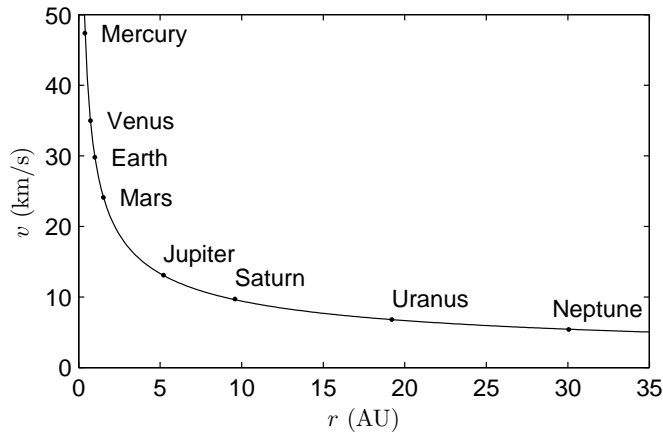}
  \caption{Orbital velocities for solar system planets as a function of distance from the sun.
  Solid line is the Newton-Kepler prediction $v = \sqrt{GM/r}$,
  where $M$ is the mass of the sun.}
  \label{fig:KeplerRotation}
\end{figure}

Consider a stationary, massive central object of mass $M$, e.g., the sun, 
orbited by an object of mass $m$ in uniform circular motion.
The radial component of Newton's second law,
$F_r = m a_r$, becomes
\begin{equation}
\label{eq:Newton}
\frac{GMm}{r^2} = m \; \frac{v^2}{r} ,
\end{equation}
where Newton's law of gravitation has been substituted on the left-hand-side,
and the centripetal acceleration formula has been substituted on the right.
This results in a version of Kepler's third law of planetary motion
that relates the velocity $v$ of the orbiting object to the radius $r$ of its circular orbit
\begin{equation}
\label{eq:K3}
v = \sqrt{ \frac{GM}{r}} .
\end{equation}
This can be used, for example, to predict the orbital velocities of the planets,
and a comparison of that prediction with observed velocities is shown in 
Fig.~\ref{fig:KeplerRotation}.

It is slightly problematic to consider this an actual prediction, however,
because Newton developed his law of gravitation specifically to agree with Kepler's third law.
In fact, after
Christiaan Huygens had solved the centripetal acceleration problem in 1658,
Edmund Halley, Christopher Wren and Robert Hooke were all able
to use this information, along with Kepler's third law, to infer that the attractive force between the Sun 
and each of the planets was an inverse square force.\cite{Henry00}
Only Newton, however, was able to solve the ``inverse problem,''
and show that an inverse square central force resulted in elliptical planetary orbits.

\subsection{The Prediction}
\label{sec:Prediction}

Today, more three hundred years later, Newton's coherent ``theory of gravitation'' 
(which includes not only the inverse-square force law but Newton's three laws of motion)
has undergone numerous tests, 
from predicting the motion of baseballs to the motion of Apollo astronauts.
However,
as discussed above, it is at best a provisional theory.
Is it applicable to the motion of entire galaxies?
At first glance, the answer appears to be ``no,''
since 
Newton's law of gravitation has not been directly confirmed at large distances
and
Newton's laws of motion have not been directly confirmed at small accelerations.
For example, the distances and accelerations in question are approximately those of the sun
as it orbits the center of the Milky Way at a radial distance of about 8.3 kpc,
and experiences a centripetal acceleration of about $2 \times 10^{-10}$ m/s$^2$.
However,
this act of applying theories beyond the regime for which they were developed,
and testing the resulting predictions,
has been a primary method of searching for universal laws.

\begin{figure}
\includegraphics[width=\figwidth]{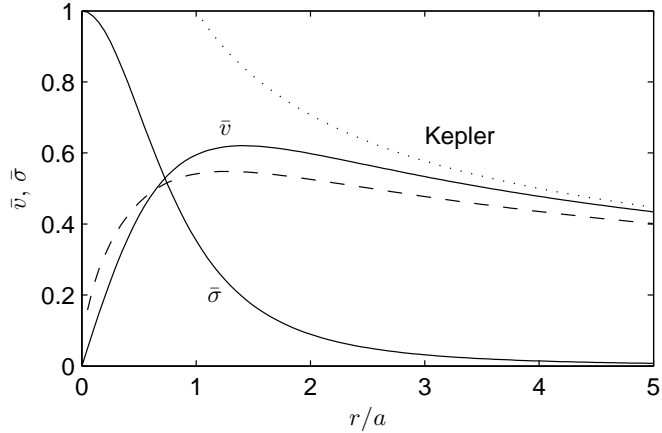}
  \caption{Scaled orbital velocity $\bar{v}$ and scaled surface density $\bar{\sigma}$ 
  for a flat disk galaxy using the Kuzmin-Toomre model given in Eqs.~(\ref{eq:surface}) and (\ref{eq:KuzminVelocity}).
  Here, $\bar{v} \equiv v \sqrt{a/GM}$ and $\bar{\sigma} \equiv \sigma (2\pi a^2/M)$.
  The dotted line is the Kepler prediction assuming the entire mass of the galaxy is concentrated at the center,
  and the dashed line is a plot of Eq.~(\ref{eq:K3mod}),
  where the integration of Eq.~(\ref{eq:surface}) gives $M(r)/M 
  = 1-a/\sqrt{r^2+a^2}$.}
  \label{fig:KuzminDisk}
\end{figure}

For spherically symmetric galaxies,
Newton's shell theorems can be used to obtain the analog to Eq.~(\ref{eq:K3})
\begin{equation}
\label{eq:K3mod}
v = \sqrt{ \frac{GM(r)}{r}} ,
\end{equation}
where $M(r)$ is the mass interior to the orbital location $r$.
For disk-type galaxies with cylindrical symmetry, however, such as spirals,
there are no such theorems.
But Poisson's equation, $\nabla^2 \Phi = 4\pi G \rho$, allows us to obtain
self-consistent functions for the gravitational potential $\Phi$ and mass density $\rho$,
from which the orbital velocity can be obtained.
Even though it may not match any particular spiral galaxy,
one simple example that has all the expected properties
is the so-called Kuzmin-Toomre model.\cite{Toomre63}
Here, the potential 
\begin{equation}
\Phi(r,z) = - \frac{GM}{\sqrt{r^2 + (|z| + a)^2}}
\end{equation}
results in an infinitesimally thin disk of surface density
\begin{equation}
\label{eq:surface}
\sigma(r,z=0) = \frac{Ma}{2 \pi (r^2 + a^2)^{3/2}} ,
\end{equation}
where the $z$-axis is the axis of symmetry of the disk,
$r=\sqrt{x^2+y^2}$ is the cylindrical coordinate in the plane of the disk,
and $M$ and $a$ are the total mass and the scale size of the galaxy, respectively.
The orbital velocity $v(r)$ of an object in circular motion in the $z=0$ plane
can be easily calculated
\begin{equation}
\label{eq:KuzminVelocity}
v^2(r) = r \left. \frac{\partial \Phi}{\partial r} \right|_{z=0} = \frac{GMr^2}{(r^2 + a^2)^{3/2}} ,
\end{equation}
which does not agree with Eq.~(\ref{eq:K3mod})
but does asymptotically approach $v \approx \sqrt{GM/r}$ when $r \gg a$.
This velocity, as well as the surface density in Eq.~(\ref{eq:surface}), is shown in 
Fig.~\ref{fig:KuzminDisk}.
Also shown are the Kepler solution for a point mass $M$ at the center, Eq.~(\ref{eq:K3}),
and the approximation of Eq.~(\ref{eq:K3mod}) using $M(r)$ calculated from Eq.~(\ref{eq:surface}).
This approximation can be calculated for any galaxy that has a known density distribution, 
and while it does not agree quantitatively with the actual orbital velocity
because the shell theorem does not hold,
it does have similar properties.

Therefore, the general features of the prediction for the rotation curve for spiral galaxies
with finite mass are the following:
low velocities near the center of the galaxy,
rising to a maximum at the ``turnover radius,''
which is the radial location where approximately half the galaxy mass is interior,
then decreasing and asymptotically approaching
the Kepler solution far from the center.

\subsection{The Observation}
\label{sec:Observation}

\begin{figure}
\includegraphics[width=\figwidth]{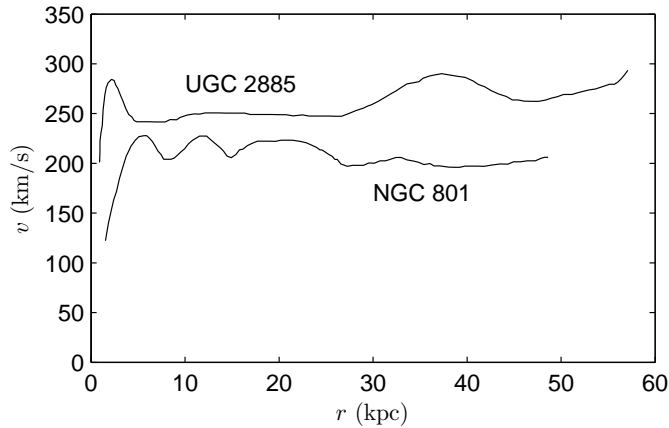}
  \caption{Velocity rotation curves for two spiral galaxies,
  UGC 2885 and NGC 801.
  After Fig.\ 5 from Rubin et al.\cite{Rubin80}
  The radial distance was obtained assuming that the Hubble constant was 50 km/s/kpc,
  and the error bars are about $\pm 25$ km/s.}
  \label{fig:Rubin80Fig5}
\end{figure}

In 1970, Rubin and Ford\cite{Rubin70}
found that the integrated mass
of the Andromeda galaxy, $M(r)$,
as inferred from the velocity rotation curve,
continued to increase out to the last measured object at $r=24$ kpc,
and that
``extrapolation beyond that distance is clearly a matter of taste.''\cite{Rubin70p394}
This meant that the total mass of the Andromeda galaxy was not measurable using their dynamical technique.
Then, Rubin et al.\cite{Rubin80} 
measured the rotation curves for 21 spiral galaxies,
two of which are shown in 
Fig.~\ref{fig:Rubin80Fig5},
and found that
``\textit{None} of the rotation curves have the classical shape,
adopted so frequently in the past,
of a long nearly Keplerian drop in velocity after the initial rapid rise.''
In fact, Fig.~\ref{fig:Rubin80Fig5}
shows that the rotation curves are flat out to large distances,
implying that the galaxies possess large amounts of mass that are not collocated
with the regions that produce light,
hence the moniker ``dark matter.''
In fact,
since the rotation curves do not turn over,
``there is no spiral galaxy with a well determined mass.''\cite{BTp599}

Taking a cue from the successful prediction of the existence of Neptune,
and assuming that all of our physical hypotheses are correct,
a reasonable
guess is that the extra mass is due to normal matter,
but in a form that does not emit light.
Initial ideas included Zwicky's list above,
or more specifically
low-mass objects that do not sustain fusion,
called brown dwarfs,
or a population of black holes and neutron stars.\cite{BTp589}
Before we look at the obvious, however,
it is worthwhile to consider the non-obvious.

\subsection{The Falsification?}

What is to be made of this result?
There is a prediction, Eq.~(\ref{eq:KuzminVelocity}),
and an observation, Fig.~\ref{fig:Rubin80Fig5},
that do not agree.
Normally physicists would insist on a quantitative agreement,
but these do not even agree qualitatively.
Following Duhem,
it is difficult to determine which hypotheses are being tested with this measurement,
because there are many that have gone into the calculation,
both explicitly and implicitly.
The implicit assumptions are probably too numerous to list,
but they include ideas such as our understanding of how stars (i.e., matter) create light.
Therefore all of our knowledge of nuclear physics and how specific nuclear reactions
occur in the hot, dense cores of stars can be brought into question.

However,
there are three explicit hypotheses that are necessary to obtain Eq.~(\ref{eq:Newton}),
namely,
Newton's second law of motion,
Newton's law of gravitation,
and 
Huygens' centripetal acceleration formula.
Because these three have been some of the primary workhorses for classical mechanics
in the 300 years since their development,
physicists are loathe to declare them incorrect.
Nevertheless,
before they are deemed unassailable,
it is worthwhile to investigate modifications that would solve this problem.

First,
almost no one advocates abandoning centripetal acceleration,
primarily because it follows from the definition of acceleration and Euclidean geometry,
and any modification would have far-reaching consequences in almost all areas of physics.
This, of course,
is the tricky part of refining a previous hypothesis:
all the phenomena that have been correctly predicted in the past must remain so.

Second,
modifying the other two laws have both been tried.
For example,
since the discrepancy arises at large distances,
if another term were added to Newton's gravitation force
\begin{equation}
F_G = mM \left( \frac{G}{r^2} + \frac{A}{r} \right) ,
\end{equation}
then when $r \gg G/A$,
the predicted rotation curve would be independent of $r$, i.e., flat,
and gravitational physics at smaller scales would be preserved.
Unfortunately, 
Aguirre\cite{Aguirre03} points out that this does not work for at least two reasons.
If the parameter $A$ is independent of the mass of the galaxy, $M$,
then the prediction for the orbital velocity at the galaxy's edge becomes $v_\infty \approx A$,
and does not agree with another observation,
namely the so-called Tully-Fisher relation.\cite{TFdetails}
If, on the other hand,
$A$ is allowed to depend on $M$ in order to satisfy the Tully-Fisher relation,
then $A \sim M^{-1/2}$,
and Newton's law of gravitation is nonlinear.
This is
a property that physicists typically only allow after other modifications have been tried.

Third,
since the discrepancy also arises at small accelerations,
Newton's second law can be modified in such a way as to leave
unchanged the high-acceleration regime in which it works so well.\cite{Milgrom83}
For example, if $F=ma$ is replaced with
\begin{equation}
\label{eq:MOND}
F = ma \left[ \frac{a}{\sqrt{a^2 + a_0^2}} \right] ,
\end{equation}
then when $a \ll a_0$, 
if the force is normal gravity, $GMm/r^2$,
then the orbital velocity is a constant,
$v^4 = GMa_0$.
(Of course, $a$ is still $v^2/r$.)
The function in square brackets in Eq.~(\ref{eq:MOND}) is just an example,
and any function of $a$ that is unity for large $a$ and proportional to $a$ for small $a$ will work.
Comparison with spiral galaxy rotation curves\cite{Milgrom02} shows that the acceleration
at which Newton's second law must be modified is $a_0 \approx 2 \times 10^{-10}$ m/s$^2$,
just the range that was anticipated when looking at a typical star like the sun,
far from the galaxy's center.
Note that the limiting velocity agrees with the Tully-Fisher relation,
as required above.

This last line of reasoning is MOND,
developed by Mordehai Milgrom in the 1980s.\cite{Milgrom83,Milgrom02,Sanders02}
Currently it is a viable alternative to dark matter,
although it is not universally accepted
because it, too, has problems.
While it predicts galaxy rotation curves well,
it does not agree with cluster observations.
Also, any accepted change that fits all the standard dark matter observations
listed above 
(clusters, rotation curves, lensing, and CMB effects)
must agree with general relativity.
Developing a general relativistic version of MOND is a current area of research.\cite{Bekenstein04}

\section{Conclusion}
\label{sec:conclusion}

The dark matter problem is still a problem.
The leading paradigm is that Newton's laws and general relativity are correct,
and a new type of particle that only interacts gravitationally
(and not electromagnetically, because then it would emit photons)
makes up this extra mass.
The same type of elimination process that I have just gone through has been used to eliminate
several types of particles, for example, neutrinos,
and we are currently left with a knowledge of what dark matter is \textit{not}.
In addition to a deeper understanding of the scientific method,
the exercise above is useful to remind students not to be afraid to challenge accepted dogma,
and to realize that what they are learning is only our current best hypothesis of how the universe works.
On the other hand,
it is equally important to remember that 
extraordinary claims require extraordinary evidence,
and that all tests of a nonstandard theory must be made before its acceptance.
Regarding dark matter,
the physics community appears to be nearing the stage expressed by 
Sherlock Holmes:
``How often have I said to you that when you have eliminated the impossible, whatever remains,
\textit{however improbable},
must be the truth?''\cite{Holmes}
But
Duhem reminds us that we will never be sure that we have eliminated everything.

\section*{Acknowledgements} I would like to thank Michael Griffith for useful discussions.


\begin{thebibliography}{}
\bibitem{TFdetails}
The Tully-Fisher relation is an empirical observation that
the a spiral galaxy's visible (i.e., baryonic) mass
is proportional to a power of the orbital velocity,
$M \propto v_\infty^\alpha$, where $\alpha$ is about 4.
This relation is a strong filter for theoretical solutions to the dark matter problem.
\end{thebibliography}
\end{document}